\documentclass[aps,prl,reprint,superscriptaddress]{revtex4-2}

\usepackage{amsmath,amssymb,bm}
\usepackage{physics}
\usepackage{graphicx}
\usepackage[colorlinks=true,linkcolor=blue,citecolor=blue,urlcolor=blue]{hyperref}

\begin{document}

\title{Local Sources of Phase Curvature and Rigidity in Finite Quantum Matter}

\author{Riccardo Castagna}
\affiliation{Consiglio Nazionale delle Ricerche (CNR), Italy\\
riccardo.castagna@cnr.it}

\date{\today}

\begin{abstract}
Finite coherent quantum systems exhibit a nontrivial response to \emph{local sources of phase curvature}, which cannot be reduced to conventional forces, disorder-induced localization, or simple gap opening.
Here we show that, in finite fermionic rings, a localized symmetry-breaking perturbation acts as a source of phase curvature in the many-body Hilbert space, inducing an anomalous breakdown of global phase rigidity.
Starting from a Hubbard--Peierls description, we derive an effective field-theoretic functional in which the inverse local susceptibility defines a phase-rigidity scale controlled by system size and electronic correlations.
This rigidity quantifies the resistance of a coherent many-body state to geometric deformation of its phase structure, rather than to energetic localization.
We demonstrate that interactions enhance phase rigidity in finite systems, counter to naive expectations based on single-particle localization, and that rigidity loss may occur without a direct correspondence to gap formation.
Molecular $\pi$-electron rings and mesoscopic quantum circuits provide experimentally accessible realizations of this regime, establishing a direct connection between local phase curvature, geometric rigidity, and coherence-driven phenomena across finite quantum matter.
\end{abstract}

\maketitle

\section{Introduction}

The response of coherent quantum systems to external perturbations often transcends the conventional force--response or energy--gap paradigms \cite{Imry2002, ButtikerImryLandauer1983}. In systems where coherence is sustained over finite geometries, localized perturbations---whether magnetic (e.g., flux insertion), electrostatic, or structural---can precipitate global rearrangements of the many-body phase structure \cite{Thouless1983, Berry1984, Giamarchi2004}. Such transformations occur even in the absence of extended disorder or macroscopic symmetry breaking, posing a fundamental challenge to our understanding of how coherence redistributes in quantum materials where the state manifold is encoded in the Berry curvature and quantum metric \cite{Nagaosa2010, Xiao2010, Provost1980}.

Canonical examples of geometric response are now well established. In the anomalous Hall effect, transverse transport arises from Berry curvature in momentum space rather than from Lorentz forces, revealing the geometric origin of an observable traditionally attributed to external fields \cite{Nagaosa2010, Haldane2004}. Similarly, in optical systems, engineered phase gradients give rise to anomalous reflection and refraction, where beam deflection emerges from geometric phase rather than from material interfaces or refractive-index discontinuities \cite{Yu2011, Bliokh2015}. In these cases, the measurable response is governed by the local geometry of the quantum state manifold, encoded in the quantum geometric tensor whose imaginary and real parts define, respectively, Berry curvature and quantum metric \cite{Provost1980, Xiao2010}.

Here we show that, in finite fermionic rings, localized symmetry-breaking perturbations act as sources of phase curvature in the many-body state manifold, leading to an anomalous breakdown of global phase rigidity governed by geometry rather than by energetic localization. We demonstrate that this phase rigidity is enhanced by electronic correlations in finite systems, counter to naive expectations based on single-particle localization, and that its loss may occur without a direct correspondence to gap formation. These results identify phase rigidity as a geometric response property of coherent many-body states and establish a fermionic counterpart to coherence-driven anomalous phenomena previously observed in optical systems.
It is worth emphasizing that the suppression of the Peierls susceptibility by repulsive interactions is well understood in the thermodynamic limit, where it is commonly interpreted as a direct consequence of charge-gap opening.
Here we demonstrate that, in finite coherent systems relevant to present quantum technologies, the same suppression acquires a qualitatively different meaning.
In this regime, the collapse of the local bond susceptibility defines an intrinsic phase-rigidity scale $\mathcal{R}(U,N)$ that does not simply follow the evolution of the charge gap $\Delta_{\rm ch}$.
While $\Delta_{\rm ch}$ quantifies energetic incompressibility, $\mathcal{R}$ measures the resistance of the many-body phase structure to localized geometric deformation.
This distinction establishes phase rigidity as an independent diagnostic of coherence protection in fermionic qubits and molecular rings, beyond gap-based energetics.

\section{Model and Effective Functional}

We consider a one-dimensional Hubbard--Peierls ring of $N$ sites at half-filling, described by $\hat{H}=\hat{H}_{\mathrm{el}}+\hat{H}_{\mathrm{lat}}$. Periodic boundary conditions ($j+1\equiv 1$) are assumed, with a magnetic flux $\Phi$ threading the ring incorporated via a Peierls phase factor $e^{i\phi/N}$, where $\phi = 2\pi \Phi/\Phi_{0}$ is the dimensionless twist angle and $\Phi_{0}=h/e$ the flux quantum. The electronic Hamiltonian reads:
\begin{equation}
    \hat{H}_{\text{el}}
    = -\sum_{j,\sigma} t_j \Bigl(e^{i\phi/N}\hat{c}^\dagger_{j\sigma}\hat{c}_{j+1\sigma} + \text{h.c.}\Bigr)
    + U \sum_{j} \hat{n}_{j\uparrow}\hat{n}_{j\downarrow},
    \label{eq:Hamiltonian}
\end{equation}
where $\hat{c}^\dagger_{j\sigma}$ creates an electron at site $j$ with spin $\sigma$, and $U$ is the on-site repulsion. The hopping amplitudes $t_j=t_0(1-\alpha \delta_j)$ are modulated by bond distortions $\delta_j$, with $\alpha$ being the electron--phonon coupling.

We treat the lattice in the adiabatic limit, where $\{\delta_j\}$ act as static classical fields subject to a restoring elastic potential:
\begin{equation}
    \hat{H}_{\text{lat}} = \frac{K}{2}\sum_{j}\delta_j^{\,2}.
    \label{eq:Hlat}
\end{equation}
This approximation isolates the geometric response of the correlated electronic ground state to \textit{localized} structural perturbations, distinct from dynamic phonon effects \cite{Heeger1988, Baeriswyl1992}.

To quantify this response, we focus on the stability of coherence against local deformation. We construct an effective field-theoretic functional for the bond order parameter field derived from $\{\delta_j\}$. We define the inverse \textit{local} susceptibility derived from this functional as the phase-rigidity scale, $\mathcal{R}$, which governs the geometric rearrangement of the many-body phase manifold under localized constraints.

To derive the phase-rigidity scale, we adopt a linear perturbative approach that is effectively equivalent to a quadratic (Gaussian) Ginzburg--Landau expansion around the uniform reference configuration ($\delta_j=0$), without assuming spontaneous long-range order in the finite ring. We introduce the bond operator conjugate to the distortion field,
\begin{equation}
\hat{B}_j=\sum_{\sigma}\Bigl(e^{i\phi/N}\hat{c}^\dagger_{j\sigma}\hat{c}_{j+1\sigma}+\text{h.c.}\Bigr),
\end{equation}
so that the Peierls modulation $t_j=t_0(1-\alpha\delta_j)$ couples $\delta_j$ to the local coherent bond (phase) structure of the many-body state.

Treating the set of localized distortions $\{\delta_j\}$ as a perturbation, the change in the total ground-state energy up to second order reads
\begin{equation}
\Delta E[\{\delta\}] \simeq \frac{1}{2}\sum_{i,j}\delta_i\,\mathcal{K}_{ij}\,\delta_j,
\label{eq:GL_functional}
\end{equation}
which defines the stiffness (inverse susceptibility) kernel
\begin{equation}
\mathcal{K}_{ij}=K\,\delta_{ij}-\alpha^2 t_0^{\,2}\,\chi_{ij}^{\text{bond}}.
\label{eq:Kernel}
\end{equation}
The first term is the bare elastic restoring cost, while the second term describes the electronic softening mediated by the \emph{static} bond--bond susceptibility. In Lehmann representation,
\begin{equation}
\chi_{ij}^{\text{bond}}
= 2\sum_{n\neq 0}\frac{\mathrm{Re}\,\langle 0|\hat{B}_i|n\rangle\langle n|\hat{B}_j|0\rangle}{E_n-E_0}
\equiv -\frac{\partial^2 E_{\mathrm{el}}}{\partial t_i\,\partial t_j},
\label{eq:Chi_bond}
\end{equation}
where $|0\rangle$ and $|n\rangle$ denote the ground and excited eigenstates of the unperturbed electronic Hamiltonian $\hat{H}_{\text{el}}(\{\delta=0\})$.

The quadratic functional in Eq.~\eqref{eq:GL_functional} defines a deformation landscape for the bond field. The stability of the coherent phase structure against a \emph{localized} perturbation at bond $j$ is controlled by the corresponding diagonal element of the stiffness kernel. We therefore define the \emph{local phase rigidity} as
\begin{equation}
\mathcal{R}\equiv \mathcal{K}_{jj}=K-\alpha^2 t_0^{\,2}\,\chi_{\mathrm{loc}},
\label{eq:Rigidity_Def}
\end{equation}
with $\chi_{\mathrm{loc}}\equiv \chi^{\text{bond}}_{jj}$. The scale $\mathcal{R}$ quantifies the geometric hardness of the many-body bond (phase) manifold: decreasing $\mathcal{R}$ signals local softening of the coherent state against bond-field curvature, and in the extreme $\mathcal{R}\to 0$ it indicates a tendency to nucleate nonuniform distortions (e.g., dimerization-like patterns) once the electronic energy gain compensates the elastic cost.

Numerically, $\chi_{\rm loc}$ can be obtained without explicit spectral sums by evaluating the curvature of the ground-state energy under a single-bond hopping modulation $t_j=t_0(1+\epsilon)$,
\begin{equation}
\chi_{\rm loc} \simeq - \frac{E_0(+\epsilon)+E_0(-\epsilon)-2E_0(0)}{(t_0\epsilon)^2},
\label{eq:NumDeriv}
\end{equation}
which only requires ground-state Lanczos calculations.

\section{Results and Discussion}

The physical significance of the rigidity scale $\mathcal{R}$ extends beyond simple energetics, as the local bond susceptibility is directly connected to the quantum geometry of the ground-state manifold.
For a local coupling parameter $\lambda\equiv t_j$, the parametric derivative of the Hamiltonian is proportional to the bond operator itself, $\partial_\lambda \hat{H} \propto \hat{B}_j$. As a result, both the static susceptibility $\chi^{\rm bond}_{jj}$ and the diagonal element of the quantum metric tensor $g_{\lambda\lambda}$ associated with local bond deformations admit Lehmann representations governed by the same transition matrix elements $\langle 0|\hat{B}_j|n\rangle$. They differ only in the spectral weighting,
\begin{align}
    \chi^{\rm bond}_{jj} &\sim \sum_{n\neq 0} \frac{|\langle 0|\hat{B}_j|n\rangle|^2}{E_n-E_0}, \\
    g_{\lambda\lambda} &\sim \sum_{n\neq 0} \frac{|\langle 0|\hat{B}_j|n\rangle|^2}{(E_n-E_0)^2}.
\end{align}
In the thermodynamic limit this behavior is commonly interpreted as Peierls suppression; in finite coherent systems the same suppression defines a geometric rigidity scale decoupled from gap formation.
While the present work focuses on the susceptibility-defined rigidity scale $\mathcal{R}$, Eq.~(10) shows that the same matrix elements control the quantum metric associated with local bond deformations.
This identifies $\mathcal{R}$ as the energetic manifestation of an underlying geometric stiffness of the many-body ground-state manifold.

An enhanced local susceptibility therefore signals a geometrically soft manifold in the space of many-body states, where the distance between neighboring ground states under local bond deformations increases. Accordingly, the rigidity scale $\mathcal{R}$ quantifies the geometric hardness of the coherent many-body state against localized sources of phase curvature, rather than its resistance to energetic localization.

\begin{figure}[t]
\includegraphics[width=\columnwidth]{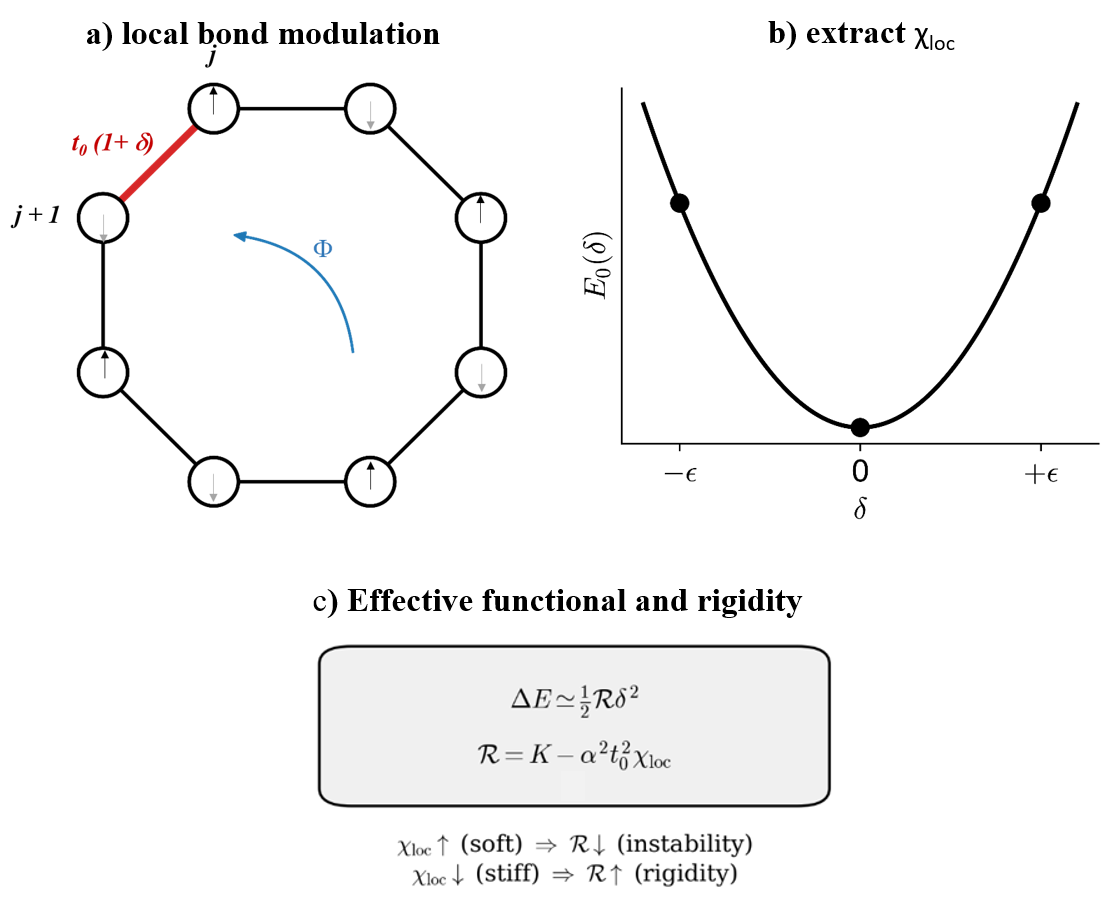} 
\caption{
\textbf{Local sources of phase curvature and definition of the rigidity scale.}
(a) A finite Hubbard--Peierls ring at half filling subject to a \emph{localized} bond modulation
$t_j=t_0(1+\delta)$, which acts as a local symmetry-breaking source.
(b) The local susceptibility $\chi_{\mathrm{loc}}$ is extracted from the curvature of the
many-body ground-state energy under the bond modulation,
$\chi_{\mathrm{loc}}\equiv -\partial_\delta^2 E_0\!\mid_{\delta=0}$.
(c) Within the quadratic effective functional
$\Delta E[\{\delta\}]\simeq \frac12\sum_{i,j}\delta_i\,\mathcal K_{ij}\,\delta_j$,
the diagonal element $\mathcal K_{jj}=K-\alpha^2 t_0^{\,2}\chi_{\mathrm{loc}}\equiv\mathcal R$
defines the local phase-rigidity scale.
}
\label{fig:schematic}
\end{figure}

To disentangle geometric rigidity from purely spectral (energetic) effects, we directly compare the local bond susceptibility $\chi_{\mathrm{loc}}(U)$ with the charge gap $\Delta_{\rm ch}(U)$ extracted from the same finite-size Hubbard rings.
The charge gap is computed in the standard way from ground-state energies in adjacent particle-number sectors,
\begin{equation}
\Delta_{\rm ch}(U;N)\equiv E_0(N+1)+E_0(N-1)-2E_0(N),
\label{eq:charge_gap_def}
\end{equation}
where $E_0(N)$ denotes the ground-state energy of $N$ electrons.

\begin{figure}[t]
\includegraphics[width=0.9\columnwidth]{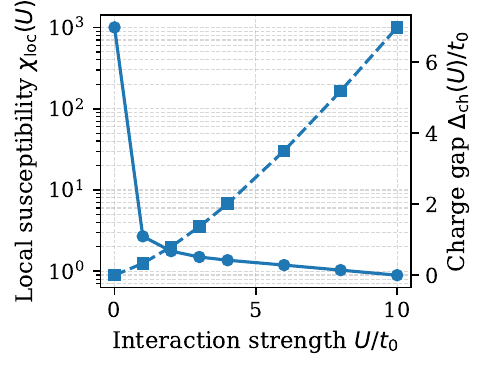} 
\caption{
\textbf {Local susceptibility does not track the charge gap.}
The local susceptibility $\chi_{\mathrm{loc}}(U)\equiv -\partial_\delta^2 E_0\!\mid_{\delta=0}$
(left axis, logarithmic scale), extracted from the curvature of the ground-state energy under a single-bond modulation,
is reported together with the charge gap
$\Delta_{\mathrm{ch}}(U)$
(right axis) for a Hubbard ring with $N=8$ sites at half filling.
While $\Delta_{\mathrm{ch}}(U)$ increases monotonically with interaction strength,
$\chi_{\mathrm{loc}}(U)$ is sharply reduced for $U>0$ and then varies only weakly (or decreases) across the correlated regime.
Thus, the local susceptibility does not track the gap, showing that coherence protection is not a trivial consequence of spectral gapping.
}
\label{fig:rigidity}
\end{figure}

Figure~\ref{fig:rigidity} reports the evolution of $\chi_{\mathrm{loc}}(U)$ and $\Delta_{\mathrm{ch}}(U)$ for a ring of $N=8$ sites.
In the noninteracting limit ($U=0$), $\chi_{\mathrm{loc}}$ is large (diverging with system size), reflecting the well-known tendency of one-dimensional fermionic systems toward Peierls-like bond softening.
Remarkably, increasing $U$ drastically suppresses $\chi_{\mathrm{loc}}$. Since the phase rigidity is defined as $\mathcal{R}=K-\alpha^{2}t_{0}^{\,2}\chi_{\mathrm{loc}}$, this suppression implies an interaction-driven enhancement of $\mathcal{R}$.

This demonstrates that electronic correlations stabilize the coherent many-body phase structure against local sources of phase curvature, effectively stiffening the quantum state manifold.
Importantly, the comparison in Fig.~\ref{fig:rigidity} reveals that the evolution of $\chi_{\mathrm{loc}}$ does not strictly mirror the opening of the charge gap. While $\Delta_{\mathrm{ch}}$ opens linearly in the Mott regime, the susceptibility collapses abruptly at weak coupling. This decoupling highlights that phase rigidity is a geometric response property controlled by local bond susceptibility rather than a direct proxy of spectral gap opening.

\section{Conclusions}
Our results suggest an experimentally accessible route to probe geometric phase rigidity in finite coherent matter. In mesoscopic quantum rings, magnetic flux control directly implements the twist $\phi$ and enables phase-sensitive measurements via persistent currents, while local gates or engineered bond disorder provide controlled symmetry-breaking perturbations. On the molecular side, $\pi$-electron nanorings and cyclic conjugated architectures offer a minimal finite realization of the same physics. In both settings, the key prediction is an interaction-driven enhancement of the local rigidity scale $\mathcal{R}$ and a rigidity breakdown not trivially reducible to gap opening, underscoring the geometric nature of coherence response in finite quantum systems. A final remark concerns the role of system size.
The phase-rigidity scale $\mathcal{R}$ introduced in this work is intrinsically a local and mesoscopic quantity, designed to characterize the response of a finite coherent system to a localized source of phase curvature.
It is therefore not intended as a bulk order parameter required to remain finite in the thermodynamic limit.
In the non-interacting case, the enhancement of the local bond susceptibility reflects the well-known Peierls instability and increases with system size, as expected.
Remarkably, the introduction of electronic correlations suppresses this enhancement already at weak coupling, and this suppression is controlled by local spectral properties rather than by the global system size.
As a result, in the correlated regime relevant to molecular rings and mesoscopic quantum circuits, the rigidity scale $\mathcal{R}(U,N)$ exhibits only weak finite-size dependence.
This identifies interaction-enhanced phase rigidity as an intrinsically mesoscopic coherence-protection mechanism, rather than a finite-size artifact or a trivial consequence of gap opening.
Whether the rigidity scale survives the thermodynamic limit or represents an intrinsically mesoscopic coherence-protection mechanism is an intriguing question left for future work. 

\begin{acknowledgments}
The author acknowledges computational assistance (AI) used during the preparation of this work.
\end{acknowledgments}

\setcounter{equation}{0}
\setcounter{figure}{0}
\setcounter{table}{0}
\setcounter{page}{1}
\renewcommand{\theequation}{S\arabic{equation}}
\renewcommand{\thefigure}{S\arabic{figure}}
\renewcommand{\bibnumfmt}[1]{[S#1]}
\renewcommand{\citenumfont}[1]{S#1}


\section{Supplemental Material}




\section{S1. Derivation of the Effective Stiffness Functional}

Here we provide the derivation of the effective quadratic functional governing the response of the system to local bond distortions, leading to the definition of the stiffness kernel $\mathcal{K}_{ij}$ used in the main text.

We start from the total Hamiltonian $\hat{H}=\hat{H}_{\mathrm{el}}+\hat{H}_{\mathrm{lat}}$. 
The lattice contribution is treated classically within the harmonic approximation,
\begin{equation}
E_{\mathrm{lat}}=\frac{K}{2}\sum_j \delta_j^2 .
\end{equation}

The electronic Hamiltonian depends on the bond distortions through the modulated hoppings
$t_j=t_0(1-\alpha\delta_j)$. 
Expanding around the uniform reference configuration $\delta_j=0$, one obtains
\begin{equation}
\hat{H}_{\mathrm{el}}(\{\delta\})
= \hat{H}_{\mathrm{el}}(0)
+ \alpha t_0 \sum_j \delta_j \hat{B}_j ,
\end{equation}
where the bond operator is
\begin{equation}
\hat{B}_j=\sum_{\sigma}\left(
e^{i\phi/N}\hat{c}^\dagger_{j\sigma}\hat{c}_{j+1\sigma}
+\text{h.c.}
\right).
\end{equation}
The positive sign follows from the combination of the hopping modulation
$t_j=t_0-\alpha t_0\delta_j$ and the overall minus sign in the kinetic Hamiltonian.

Treating the distortion field as a perturbation,
$\hat{V}=\alpha t_0\sum_j\delta_j\hat{B}_j$,
standard second-order perturbation theory yields
\begin{equation}
E_{\mathrm{el}}(\{\delta\})
\simeq
E_0(0)
-\sum_{n\neq0}
\frac{|\bra{0}\hat{V}\ket{n}|^2}{E_n-E_0},
\end{equation}
where the linear term vanishes for the uniform ground state in the absence of spontaneous dimerization.

Explicitly,
\begin{align}
\Delta E_{\mathrm{el}}^{(2)}
&=
-(\alpha t_0)^2
\sum_{n\neq0}\sum_{i,j}
\delta_i\delta_j
\frac{\bra{0}\hat{B}_i\ket{n}\bra{n}\hat{B}_j\ket{0}}{E_n-E_0}
\nonumber\\
&=
-\frac{1}{2}\alpha^2 t_0^2
\sum_{i,j}
\delta_i
\chi^{\mathrm{bond}}_{ij}
\delta_j ,
\end{align}
where the factor $1/2$ accounts for the symmetrization of the quadratic form, and
$\chi^{\mathrm{bond}}_{ij}$ is the static bond--bond susceptibility in Lehmann representation.

Adding the elastic contribution, the total energy variation reads
\begin{equation}
\Delta E_{\mathrm{tot}}
=
\frac{1}{2}
\sum_{i,j}
\delta_i
\left(
K\delta_{ij}
-\alpha^2 t_0^2 \chi^{\mathrm{bond}}_{ij}
\right)
\delta_j ,
\end{equation}
which identifies the stiffness kernel
$\mathcal{K}_{ij}$ introduced in the main text.

\section{S2. Numerical Implementation}

The numerical results were obtained using Exact Diagonalization based on the Lanczos algorithm.
We consider a one-dimensional ring of $N$ sites at half filling ($N_e=N$) in the $S^z=0$ sector.
For the largest system sizes used for consistency checks ($N=10$), the Hilbert space dimension
$\binom{10}{5}^2=63\,504$ is readily handled using sparse-matrix techniques.

The local susceptibility
$\chi_{\mathrm{loc}}\equiv\chi^{\mathrm{bond}}_{jj}$
is evaluated as the curvature of the ground-state energy under a modulation of a single bond,
$t_j=t_0(1+\epsilon)$.
A symmetric finite-difference formula is employed,
\begin{equation}
\chi_{\mathrm{loc}}
\simeq
-\frac{E_0(+\epsilon)+E_0(-\epsilon)-2E_0(0)}{(t_0\epsilon)^2},
\end{equation}
with $\epsilon=10^{-3}$.
The stability of the result was verified for
$\epsilon\in[10^{-4},10^{-2}]$.
Ground-state energies were obtained with a sparse Lanczos eigensolver using a convergence tolerance of $10^{-12}$ to ensure numerical accuracy.

\section{S3. Non-Interacting Limit}

In the non-interacting limit ($U=0$), the model reduces to a tight-binding ring.
The Peierls instability criterion predicts a logarithmic divergence of the bond susceptibility with system size in the thermodynamic limit.
For finite rings, this divergence is cut off by the finite-size gap
$\Delta_{\mathrm{FS}}\sim t_0/N$.

Accordingly, $\chi_{\mathrm{loc}}$ assumes a large but finite value controlled by $N$.
For the system sizes considered in the main text, we find
$\chi_{\mathrm{loc}}\sim10^{3}$ (in units of $1/t_0$),
consistent with the pronounced softness of the free Fermi gas toward bond deformations.
The introduction of a finite interaction $U$ immediately suppresses this susceptibility, as discussed in the main text.

\end{document}